\documentclass[aps,twocolumn,nofootinbib,preprintnumbers,superscriptaddress,citeautoscript]{revtex4-1}
\pdfoutput=1
\usepackage{graphicx}
\usepackage{epsfig,amsmath,amsthm,amsfonts,mathrsfs,amssymb}
\usepackage[usenames,dvipsnames,table]{xcolor}
\usepackage{hyperref}

\hypersetup{
    pdfstartview={FitH},    
    pdftitle={},    
    pdfauthor={Eric Perlmutter, Mukund Rangamani, Massimiliano Rota},     
    colorlinks=true,       
    linkcolor=blue,          
    citecolor=red,        
    filecolor=magenta,      
    urlcolor=blue           
}

\def\Tr{{\rm Tr}}
 
\def\ket#1{\mid \! #1\rangle} 
\def\bra#1{\langle \, #1 \! \mid}

\def\ptr#1{{{#1}^\Gamma}}
\def\Eneg{{\mathscr E}}

\def\Eneg{{\mathscr E}}

\def\regA{{\cal A}} 
\def\regAc{{\cal A}^c} 
\def\rhoA{{\rho_{\regA}}} 
 
\def\entsurf{
\partial \regA}

\def\Cu#1{C_\text{univ}[{#1}]}
\def\sE{{\sf E}}
\def\Su{S_\text{u}}

\def\hX{{\widehat {\cal X}}}

\definecolor{rust}{rgb}{0.8,0.2,0.2}
\definecolor{green}{rgb}{0.1,0.8,0.2}
\definecolor{myblue}{rgb}{0.12,0.51,0.88}

\begin{document}
\title{Positivity, negativity, and entanglement}

\preprint{DCPT-15/33}

\author{Eric Perlmutter}
\email{perl@princeton.edu}
\affiliation{Department of Physics, Princeton University, Princeton, NJ 08544, USA
}

\author{Mukund Rangamani}
\email{mukund.rangamani@durham.ac.uk}
\affiliation{Centre for Particle Theory \& Department of
Mathematical Sciences, Durham University, South Road, Durham DH1 3LE, United Kingdom}

\author{Massimiliano Rota}
\email{massimiliano.rota@durham.ac.uk}
\affiliation{Centre for Particle Theory \& Department of
Mathematical Sciences, Durham University, South Road, Durham DH1 3LE, United Kingdom}

\begin{abstract}{
We explore properties of the universal terms in the entanglement entropy and logarithmic negativity in 4d CFTs, 
aiming to clarify the ways in which they behave like the analogous entanglement measures in quantum mechanics. We show that, unlike entanglement entropy in finite-dimensional systems, the sign of the universal part of entanglement entropy is indeterminate. In particular, if and only if the central charges obey $a>c$, the entanglement across certain classes of entangling surfaces can become arbitrarily negative, depending on the geometry and topology of the surface. The negative contribution is proportional to the product of $a-c$ and the genus of the surface. Similarly, we show that in $a>c$ theories, the logarithmic negativity does not always exceed the entanglement entropy.
}
\end{abstract}

\maketitle

\section{Introduction}
\label{sec:intro}

States of a quantum mechanical system are distinguished by the presence of entanglement.
Oftentimes one characterizes this by bipartitioning the system and computing the entanglement entropy, $S$. In continuum quantum systems, the natural subdivision is geometric: we partition the state across a fiducial {\em entangling surface}. While (for pure states) $S$ is the best measure of the total amount of quantum entanglement between a region and its complement, other measures provide additional information about the pattern of entanglement for the same bipartition. A natural question is: given a fixed state of the system, how does entanglement depend on the geometry and topology of the entangling surface?

While $S$ is plagued with UV divergences in a continuum QFT, its universal piece contains non-trivial physical information, including central charges and RG monotones \cite{Solodukhin:2008dh, Myers:2010tj, Casini:2011kv, Jafferis:2011zi}. In many respects, these universal terms are the natural counterparts of quantum-mechanical entropies, which are positive. Another interesting measure is the logarithmic negativity $\Eneg$  \cite{Vidal:2002zz,Audenaert:2003aa,Plenio:2005aa}, which gives an upper bound on distillable entanglement in quantum mechanics, and is thus strictly greater than the entanglement entropy. 

These intuitive analogies with quantum mechanics suggest that, in QFT, the universal, cutoff-independent terms of $S$ and of $\Eneg-S$ are also positive-definite. Indeed, this appears to be true for spherical entangling surfaces in vacuum states of CFTs in flat spacetime \cite{Myers:2010tj, Casini:2011kv, Jafferis:2011zi, Rangamani:2014ywa}. As we will prove, however, these signs depend non-trivially on the topology of the entangling surface and, in particular, can be negative. 

We focus on connected entangling surfaces in 4d CFTs, which are Riemann surfaces. While for simple topologies the universal terms are positive-definite, we show that one can always pick complex enough entangling surfaces to violate this bound. Curiously, the violation hinges on the difference of the central charges $a$ and $c$. Specifying to entanglement entropy, the universal part of $S$ becomes negative for a suitable choice of surface if and only if $a>c$; the effect is linear in the product of $a-c$ and the genus of the surface, exhibiting a novel interplay between central charges and topological sensitivity of entanglement.

\section{Entanglement measures}
\label{sec:emeasures}

Consider a (relativistic) QFT on a $d$-dimensional spacetime ${\cal B}$; the state $\rho$ (=$\ket{\psi}\bra{\psi}$ if pure) is defined on a spatial Cauchy slice $\Sigma$ at fixed time. The biparitioning is provided by geometrically dividing $\Sigma =\regA \cup \regAc$ across a smooth spacetime codimension-2 entangling surface $\entsurf$.  Defining the reduced density matrix $\rhoA =  \Tr_{\regAc}\left(\rho\right)$, the entanglement and R\'enyi entropies are:
\begin{equation}
\begin{split}
S(\rhoA) &= -\Tr\left(\rhoA\, \log \rhoA\right)  = \lim_{q\to 1}\, S^{(q)}(\rhoA)  \,,\\ 
 S^{(q)}(\rhoA) &= \frac{1}{1-q}\, \log \Tr \left(\rhoA^q\right) .
\end{split}
\label{eq:srdef}
\end{equation}	

Another quantity of interest to us will be the negativity which is defined in terms of an auxiliary  partial transposed density matrix $\ptr{\rho}$. Picking a  basis, $\ket{{\mathfrak r}_i}$ for $\regA$ and 
$\ket{{\mathfrak l}_n}$ for $\regAc$, one defines the map $\rho \to \ptr{\rho}$ as:
\begin{equation}
\bra{{\mathfrak r}_i\, {\mathfrak l}_n}\ptr{\rho}\ket{{\mathfrak r}_j 
{\mathfrak l}_m}=
\bra{{\mathfrak r}_i\, {\mathfrak l}_m} \rho \ket{{\mathfrak r}_j 
{\mathfrak l}_n} \,.
\label{eq:ptrdef}
\end{equation}  
Thence, the \textit{logarithmic negativity}  is given in terms of the trace norm, $\| {\cal O}\|$, viz.,
\begin{equation}
\Eneg(\rho)=\log{\|\ptr{\rho}\|} = \log\left[\Tr\left(\sqrt{(\ptr{\rho})^\dagger \, \ptr{\rho}}\right)\right]
\end{equation}
It is important to note that the negativity provides an upper bound on entanglement inherent in the state and as such satisfies $\Eneg \geq S_\regA$.
For mixed states the negativity is notoriously hard to compute (see \cite{Calabrese:2012ew,Calabrese:2012nk,Kulaxizi:2014nma,Calabrese:2014yza} for results in 2d CFTs). For pure states one can relate it to the R\'enyi entropy \cite{Calabrese:2012ew}, viz., $\Eneg(\rho=\ket{\psi}\bra{\psi}) = S^{(\frac{1}{2})} (\rhoA)$. This was exploited in \cite{Rangamani:2014ywa} to explore negativity for ball-shaped regions with $\entsurf= {\bf S}^{d-2}$ in a CFT vacuum.

Local dynamics of a QFT implies that the measures, collectively denoted as $\sE=\{S, S^{(q)}, \Eneg\} $ are UV divergent. Given a UV cut-off $\epsilon$ one finds \cite{Ryu:2006ef} 
\begin{equation}
\!\! \sE = \sum_{k=0}^{d-4}\, \frac{\sE_k}{\epsilon^{d-2-2k}} -
\begin{cases}
& \hspace{-3mm} \Cu{\sE}\, \log \frac{\ell_\regA}{\epsilon} +C_0 \,, \;\; d = \text{even}\\
& \hspace{-3mm} (-1)^{d-1} \, \Cu{\sE} \,,\qquad   \; d = \text{odd}
\end{cases}
\label{eq:Seps}
\end{equation}	
The leading UV-divergence obeys an area law, $\sE_0 \propto \text{Area}(\entsurf)$,  followed by scheme-dependent (but state independent) subleading pieces $\sE_k$. $\Cu{\sE}$  depends on the state and captures important universal physical information; for $\entsurf= {\bf S}^{d-2}$ in the vacuum, for instance, $\Cu{S(\rhoA)}$ is a measure of degrees of freedom.

\section{Entangling geometries}
\label{sec:entgeo}

Our specific interest will be in $d=4$, where $\entsurf$ can be taken to be a Riemann surface of arbitrary topology; we will explore how topology imprints itself on the entanglement. Two particular issues will be of concern to us:
\begin{itemize}
\item Is $\Cu{S(\rhoA)} \equiv \Su$ sign-definite?

\item Consider the ratio 
\begin{equation}
{\cal X} = \frac{\Cu{\Eneg(\rho)}}{\Cu{S(\rhoA)}}  
\label{eq:Xdef}
\end{equation}	
defined originally in \cite{Rangamani:2014ywa}. Is $ {\cal X}-1 \equiv \hX$ positive definite?
\end{itemize}

Recently, variants of this question have been addressed by several authors: \cite{Allais:2014ata} and \cite{Mezei:2014zla} examined the shape dependence of entanglement entropy for entangling surfaces of spherical topology in $d$ dimensions. The latter conjectured that $\entsurf={\bf S}^{d-2}$ minimizes the universal term in that topological class. In \cite{Astaneh:2014uba}, the authors searched for surfaces that maximize entanglement entropy keeping the area of $\entsurf$ fixed. They related the construction to a well-known geometric problem called the Willmore conjecture \cite{Willmore:1965uq}. Their conclusion was that in $d=4$, the maximizer over {\it all} topological classes is $\entsurf={\bf S}^2$. We will make use of their techniques to show that this is, in fact, not true for general CFTs, and appears to rely on the tacit assumption that $a=c$.

\subsection{Universal R\'enyi entropy of 4d CFTs}
\label{sec:ren4d}

To make progress we will make use of a result for $\Cu{S^{(q)}} \equiv \Su(q)$ in 4d CFTs 
\cite{Fursaev:2012mp} (nb $\Su = \Su(1)) $:
\begin{equation}
\Su(q) = \frac{f_a(q)}{2\pi}\, {\cal R}_{\entsurf} + \frac{f_b(q)}{2\pi}\, {\cal K}_{\entsurf }- \frac{f_c(q)}{2\pi}\, {\cal C}_{\entsurf}
\label{eq:sq}
\end{equation}	
The geometric quantities depend on intrinsic and extrinsic geometry of $\entsurf \subset {\cal B}$. For an embedded 2-surface $X$,
\begin{equation}
\begin{split}
{\cal R}_X &= \int_X d^{2}x \,\sqrt{\gamma}\; ^{\gamma}\!R \,, \\
 {\cal K}_X & = \int_X  d^{2}x \,\sqrt{\gamma} \left[ K^{\alpha}_{ij}\, K^{\alpha ij} - \frac{1}{2}\, (K^{\alpha\, i}_i)^2  \right]\\
 {\cal C}_X & = 2\int_X  d^{2}x \, \sqrt{\gamma}\; C_{\mu\nu\rho\sigma}t^{\mu}s^{\nu}t^{\rho}s^{\sigma}\,
\end{split}
\label{eq:curv}
\end{equation}	
Here $\gamma_{ij} $  is the intrinsic metric on $X$, $g_{\mu\nu}$ that of the full spacetime ${\cal B}$, $K_{ij}^\alpha$ is the extrinsic curvature of $X$ with $\alpha =\{t,s\}$ indexing the two normal directions (one timelike 
$t^\mu$ and the other spacelike $s^\mu$) and ${\cal C}_X$ is the pullback of the Weyl tensor  $C_{\mu\nu\rho\sigma}$ onto $X$.

We see here a clean separation between the geometric data and the intrinsic field theory features captured by the coefficient functions $f_{a,b,c}(q)$. In the $q\rightarrow 1$ entanglement limit \cite{Solodukhin:2008dh},
\begin{equation}
f_a(1)=a \,, \qquad f_b(1) = f_c(1) = c\,. 
\label{eq:acS}
\end{equation}	
For generic $q$, these functions are known not to obey a universal form.  

We now have some ammunition to tackle the questions we raised. For simplicity we will take ${\cal B} = {\mathbb R}^{3,1}$ (or equivalently ${\cal B} = {\bf S}^3 \times {\mathbb R}$ as appropriate for radial quantization). These backgrounds being conformally flat, one finds no contribution from $f_c(q)$, for ${\cal C}_{\entsurf} = 0$. If we further restrict attention to regions $\regA$ which lie on constant time slices, $K^t_{\mu\nu} = 0$. We can then focus on the purely spatial geometry of 2-surfaces $\entsurf$ embedded in either ${\mathbb R}^3 $  or ${\bf S}^3$. This allows us to use some useful results in Riemannian geometry to make precise statements.

With this understanding let us focus attention on $\Su(q)$ and $\hX$, and ask if they obey any sign-definiteness properties.

\subsection{Of central charges and R\'enyi coefficients}
\label{sec:}

Let us start by noting some basic results that hold for unitary CFTs. The central charges $a,c$ are positive definite and their ratio is bounded as \cite{Hofman:2008ar}
\begin{equation}
\frac{1}{3} \leq \frac{a}{c} \leq \frac{31}{18} \,.
\label{eq:acbdd}
\end{equation}	
The bounds are tighter in superconformal field theories. Recently it has been argued that the R\'enyi coefficient functions are not independent and satisfy
\begin{equation}
f_b(q) =f_c(q) = \frac{q}{q-1} \left[ a - f_a(q) - (q-1)\, f_a'(q)\right]
\label{eq:lprels}
\end{equation}	
The first of these equalities has not been shown in full generality but holds in both free and holographic CFTs \cite{Lee:2014xwa}. We will however assume this in what follows. The second has been proved directly in R\'enyi index perturbation theory \cite{Lewkowycz:2014jia}. One can further prove
\begin{equation}
f_a(q) > 0 \,, \;\; f_a'(q) < 0 \quad \Longrightarrow \quad f_c(q) > 0 \,, \quad \forall~ q
\label{eq:facbdd}
\end{equation}	
where we used \eqref{eq:lprels} in obtaining the implication. The inequalities on $f_a(q)$ follow from the fact that $S(\rhoA)$ obeys general inequalities for any ${\cal A}$ \cite{zycz}, and that  $S(\rhoA) \propto f_a(q)$ for $\entsurf={\bf S}^2$.

\subsection{Geometry of entangling surfaces}
\label{sec:}

To make progress we need to examine the geometry of the Riemann surface $\entsurf$. The intrinsic curvature contribution in \eqref{eq:sq} is topological; for compact $X$ the Gauss-Bonnet theorem relates it to the Euler number
\begin{equation}
{\cal R}_{X} = 8\pi (1-g) \,.
\label{eq:euler}
\end{equation}	
The extrinsic contribution can be noted to be positive definite (using $K^t_{ij} =0$, $K^s_{ij} = K_{ij}$)
\begin{equation}
{\cal K}_{X} = \int_{X}d^2x \, \sqrt{\gamma}\; \left( K_{ij}- \frac{1}{2}\, \gamma_{ij}\, \gamma^{kl}\, K_{kl}\right)^2
\label{}
\end{equation}	
This by itself is not sufficient, but we can invoke some geometry, see \cite{Astaneh:2014uba}. Introduce the Willmore energy functional \cite{Willmore:1965uq}\footnote{For surfaces embedded in ${\mathbb R}^3$ we can drop the contribution from the area of the surface. }
\begin{equation}
{\cal W}_{X \subset {\bf S}^3} = \frac{1}{4}\, \int_X d^2x\, \sqrt{\gamma} \left(1+ \frac{1}{4}\, (\gamma^{ij}\,K_{ij})^2\right) .
\label{}
\end{equation}	
This functional was introduced by Willmore, who explored surfaces which minimize their mean curvature. It obeys ${\cal W}_X \geq 4\pi$ for all $X$, and is minimized by the equatorial 
${\bf S}^2 \subset {\bf S}^3$. Willmore conjectured that at $g=1$, the Willmore functional obeys ${\cal W}_X \geq 2\pi^2$. This result was proven recently \cite{Marques:2014ul}; the unique minimizer is the Clifford torus (ratio between the radii being $\sqrt{2}$). This conjecture was generalized to higher genus, where there exist so-called Lawson surfaces \cite{Lawson:1970bh} $L_g$ for $g\geq 2$ satisfying
\begin{equation}
4\pi \leq {\cal W}_{L_g}  \leq 8\pi~,
\label{eq:lawson}
\end{equation}	
which are conjectured to be the unique minimizers of ${\cal W}_X$ \cite{kusner1989}. The precise value of ${\cal W}_{L_g}$ is unknown, but at every genus it has been proven that there is a surface that obeys \eqref{eq:lawson}, irrespective of being the minimizer \cite{simon, kusner1989}. These results will suffice for our purposes.\footnote{Lawson surfaces tend to be bulgy with small handles, especially as $g$ increases. We encourage the reader to peruse the numerically-constructed surfaces in Table 1 of \cite{Joshi07c.:energy} or Figure 1 of \cite{heller}.} See \cite{Brendle} for further details. 

To make use of the bounds on the Willmore functional, we exploit the Gauss-Codazzi equations, which are geometric identities which relate intrinsic and extrinsic curvatures. The relation we need is simple (cf., \cite{Astaneh:2014uba}):
\begin{equation}
{\cal W}_X  = \frac{1}{2}\, \left({\cal R}_X  + {\cal K}_X  \right)
\label{eq:gauss}
\end{equation}	
For compact $X$ we are immediately in business, since we can use the topological constraint on the Euler number and the geometric constraint \eqref{eq:lawson} of Lawson surfaces to examine bounds on $\Su$ and $\hX$. In particular, plugging \eqref{eq:gauss} into \eqref{eq:sq}, we dial up the genus, driving ${\cal R}_{\entsurf}$ negative, while restricting to Lawson surfaces $\entsurf=L_g$ which have ${\cal W}_{L_g}$ bounded from above.

\section{Entanglement bounds}
\label{sec:}
Let us begin by studying the bounds on the universal part of entanglement entropy. It is useful to treat the genus $g=0$ cases first and then consider $g\geq 1$. For a spherical entangling surface, it is known from \cite{Casini:2011kv}  that the R\'enyi entropies are related to thermal entropies on the hyperbolic cylinder $\mathbb{H}^3 \times {\mathbb R}$. The geometry is such that the extrinsic curvature term ${\cal K}_{\entsurf}$ vanishes and so $\Su(q) = 4f_a(q)$, which we have shown is positive definite, cf., 
\eqref{eq:facbdd}. It then follows as described in \cite{Rangamani:2014ywa} that $\hX = \frac{1}{a}\, f_a(\frac{1}{2}) -1$ is also positive definite. Assuming the sphere is the minimizer of $\hX$ at $g=0$, this establishes positivity for all $g=0$ entangling surfaces. Alternatively, positivity follows from \eqref{eq:Xhat} if one assumes that $\alpha_W>0$ for all CFTs. As we will discuss, this is true in all known examples. 

Let us turn to entangling surfaces with non-trivial topology. Simplifying \eqref{eq:sq} using \eqref{eq:gauss},
\begin{equation}
\begin{split}
\Su &= \frac{c}{2\pi} \left(2\, {\cal W}_{\entsurf} + \left(\frac{a}{c} -1\right) \, {\cal R}_{\entsurf} \right)  \\ 
&= \frac{c}{2\pi} \left(\frac{2a}{c}\, {\cal W}_{\entsurf} + \left(1-\frac{a}{c} \right) \, {\cal K}_{\entsurf} \right). 
\end{split}
\label{eq:suac}
\end{equation}	
These equations make it clear that there is a curious interplay between the sign of the central charge difference $c-a$, the topology and geometry of $\entsurf$, and the sign of $\Su$. While there is no constraint from toroidal topology (as $c> 0, {\cal W} >0$), we can infer that for $g\geq 2$:
\begin{itemize}
\item $a \leq c \Longrightarrow \Su >0$, $\forall\; \entsurf$ owing to the lower bound on the Willmore functional
and  positivity of ${\cal K}$.
\item $a>c  \Longrightarrow \Su \gtrless 0$.  The indefinite sign owes its origin to the fact that there are  Lawson surfaces which have  genus-independent bounded ${\cal W}$ \eqref{eq:lawson}, but ${\cal R}$ that can be made arbitrarily negative by ramping up the genus. The sign flip of $\Su$ across such surfaces occurs at a critical genus
\begin{equation}
g_c = 1+{{\cal W}_{L_{g_c}}\over 4\pi}{c\over a-c}~.
\label{gcrit}
\end{equation}
We note in passing that it is strongly believed that ${\cal W}_{L_g}$ monotonically increases with $g$ \cite{kuhnelpink}. 
\end{itemize}
In \cite{Astaneh:2014uba} it has been conjectured that $\entsurf = {\bf S}^2$  minimizes $\Su$ (assuming $a=c$). We now see that when $a> c$, there is no minimizer: $\Su$ is unbounded from below.

Note that not all higher genus surfaces will render $\Su<0$. However, this is guaranteed to occur above some critical genus for all surfaces whose Willmore energy grows slower than linearly in $g$. There are likely other families of surfaces besides the Lawson surfaces, as well as isolated surfaces that exist for particular values of $g$, that satisfy this criterion. For example, one can smoothly deform Lawson surfaces with fixed topology.\footnote{See \cite{heller} for such a construction at $g=2$, especially Figures 2 and 5 therein.}

Strictly speaking, the results above pertain to bounded regions $\regA$ so that $\entsurf$ is compact. For non-compact entangling surfaces we are not aware of any obvious upper bound on ${\cal W}$.

Let us now turn to the negativity and consider the quantity $\hX$ which was conjectured in \cite{Rangamani:2014ywa} to be positive definite. Using the definition in terms of the Willmore functional and the expressions \eqref{eq:lprels} we can write:
\begin{equation}
\begin{split}
\hX &= \frac{\alpha_R\, {\cal R}_{\entsurf} + 2\,\alpha_W {\cal W}_{\entsurf}}{2c\, \, {\cal W}_{\entsurf} + (a-c) \, {\cal R}_{\entsurf}} \\
\alpha_R &= \frac{1}{2}\, f_a'(\frac{1}{2}) +c  \,, \qquad \alpha_W =   f_c(\frac{1}{2}) - c
\end{split}
\label{eq:Xhat}
\end{equation}	
We can infer that the sign of $\hX$ depends on the coefficients $\alpha_R$ and $\alpha_W$ in a non-trivial fashion.
\begin{itemize}
\item For a toroidal entangling surface, $\hX \propto \alpha_W$ and so positivity requires $f_c(\frac{1}{2} )> c$. This is seen to be true in all known examples. 
\item At higher genus, if $a\leq c$ we require that $\alpha_R \leq 0 $ to ensure $\hX \geq 0$. 
\item On the other hand if $a>c$, we can easily end up with negative values of $\hX$: even if $\alpha_R \leq 0 $ there is some genus $g$ for which $\hX \leq 0$. This is because whilst the numerator is ensured to be positive, the denominator can be made arbitrarily negative by picking an appropriate Lawson surface. The situation cannot be remedied by changing the sign of $\alpha_R$ in any obvious manner. 
\end{itemize}

\section{Examples}
\label{sec:examples}

We have derived above some general conditions for the positivity of $\Su$ and $\hX$ in terms of the central charges. In Table \ref{tab:fft} we provide explicit results for  a class of free and holographic CFTs \cite{Hung:2011nu, Fursaev:2012mp, Lewkowycz:2014jia}.

 \begin{widetext}
\begin{center}
\rowcolors{2}{red!5}{blue!5}
\begin{table}[h]
\begin{tabular}{||c||c||c|c|c|c||c||c||}
\hline
Theory&  
	$\frac{a}{c}$ & 
	$f_a(q)$ & 
	$f_c(q)$ & 
	$\alpha_R$ & 
	$\alpha_W$ & 
	$\Su$ &
	$\hX$
 \\  \hline 
 Scalar  &
 	$\frac{1}{3}$ & 
 	$\frac{(1+q)(1+q^2)}{4\,q^3}\, a$ & 
 	 $3\, f_a(q)$ & $-{11\over 2}\,a$ & 
 	 $\frac{33}{4}\, a$ & 
 	 $\frac{(3\, {\cal W}- {\cal R})}{\pi}a$ &
 	 $\frac{11}{4}$ 
 \\ \hline 
 Fermion & 
 	$\frac{11}{18}$ & 
	$\frac{(1+q)(7+37q^2)}{88\,q^3}\, a$ &  
	$\frac{3(1+q)(7+17q^2)}{88\,q^3}\,a$ & 
	$-\frac{7}{4}\,a$ & $\frac{261}{88} \,a$  & 
	$\frac{(36\, {\cal W}-7\, {\cal R})}{22\pi}a$ &
	$\frac{77 \, {\cal R} - 261\, {\cal W}}{28\, {\cal R}- 144 \, {\cal W}}$ 
\\ \hline
 Vector & 
 	$\frac{31}{18}$ &
 	$\frac{1+q+31q^2+91q^3}{124\,q^3}\, a$ &  
 	$\frac{3(1+q)(1+11q^2)}{124\,q^3}a$ & 
 	$-\frac{11}{62}\,a$ & 
 	$\frac{63}{124}\, a$ & 
 	$\frac{(13 \, {\cal R}+ 36\, {\cal W})}{62\pi}a$ &
 	$\frac{-11{\cal R}+ 63\,{\cal W}}{26\,{\cal R}+72 \, {\cal W}}$ 
 \\ \hline 
Free ${\cal N}=4$ & 
 	$1$ & 
 	${1+q+7q^2+15q^3\over 24q^3}a$ &  
 	${(1+q)(1+3q^2)\over 8q^3}a$ & 
 	$-{11\over 12}a$ & 
 	${13\over 8}a$ &
 	${{\cal W}\over \pi}a$ &
 	${-11{\cal R}+39{\cal W}\over 24{\cal W}}$ 
 \\  \hline	
  Einstein & 
 	$1$ & 
 	${q\over 2(q-1)}(2-x_q^2-x_q^4)a$ &  
 	${3q\over 2(q-1)}(x_q^2-x_q^4)a$ & 
 	$-{3\over 4}a$ & 
 	${1+6\sqrt{3}\over 8}a$ &
 	${{\cal W}\over \pi}a$ &
 	${-3{\cal R}+(1+6\sqrt{3}){\cal W}\over 8{\cal W}}$ 
 \\  \hline	
\end{tabular}
\caption{Results for the universal terms in  R\'enyi entropy and their implications for $\Su$ and $\hX$ in a class of CFTs. We have chosen to write the answers in terms of the $a$ central charge. In the last line, we have defined $x_q\equiv {1\over 4q}(1+\sqrt{1+8q^2})$.}
\label{tab:fft}
\end{table}
\end{center}
 \end{widetext}

Several comments are in order. First, all known examples obey the inequalities $\alpha_W>0$ and $\alpha_R < 0$. We believe that these are likely to be true for all CFTs.  

Second, $\hX$ is shape-independent for the free scalar. It follows from \cite{Lewkowycz:2014jia} that $\hX$ is shape-independent only for theories whose $f_a(q)$ equals that of a free scalar; besides the scalar itself, there are no known examples of such theories. 

Finally, the free vector field is the only theory in this table with $a>c$, and indeed, we see that both $\Su$ and $\hX$ become negative for sufficiently negative ${\cal R}$ and upper-bounded ${\cal W}$, as happens for Lawson surfaces. Assuming monotonicity of ${\cal W}_{L_g}$ as a function of $g$, the critical genus is $g_c=4$. In arriving at this conclusion, we are assuming that the modular Hamiltonian that defines $f_a(q)$ includes the effects of the edge modes described in \cite{Donnelly:2014fua} and \cite{Huang:2014pfa}. This is necessary for $\Su$ to be determined by the $a$ central charge for spherical entangling regions. Curiously, ignoring these modes leads to $\Su$ being determined by $\hat{a} = \frac{16}{31}\, a $  \cite{Dowker:2010bu} which satisfies $\hat {a} < c$. Exploring the  dependence of $\Su$  on the entangling surface should reveal whether it is controlled by $\hat a$ as opposed to the physical central charge $a$; our diagnostic would simply involve a sign check for a  $g=5$ Lawson entangling surface.

\section{Discussion}
We have found that in CFTs with $a>c$, the universal term in entanglement entropy, $\Su$, necessarily becomes negative for certain higher genus entangling surfaces. The negativity ratio $\hX$ also generically becomes negative for $a>c$; if $\alpha_R\leq 0$ for all CFTs, this can only happen for $a> c$. It would be nice to establish whether $\alpha_R\leq 0$ and $\alpha_W>0$ identically, as suggested by all examples. 

Aside from the free vector, theories with $a>c$ include the IR fixed point of the $SU(2)$ model of \cite{Intriligator:1994rx}, as well as the non-Lagrangian Gaiotto-type $T_N$ theories \cite{Gaiotto:2009we}. The latter are IR limits of worldvolume theories of $N$ $M5$-branes wrapping genus-$\hat g$ Riemann surfaces. A characteristic example is the $A_{N-1}$ theory preserving ${\cal N}=2$ SUSY, which has $24(a-c) = (N-1)(\hat g-1)$ for $\hat g>1$. Central charges for a larger family of related ${\cal N}=1$ theories with $a>c$ are given in \cite{Bah:2012dg}. At large $N$ \cite{Gaiotto:2009gz}, where $a,c \propto N^3$, there is an interesting relation between $N$ and the entangling surface topology: namely, the critical genus $g_c$ in \eqref{gcrit} scales like $N^2$. The growth of $g_c$ with large $N$ will be true of any holographic theory with a sensible derivative expansion in the bulk \cite{Buchel:2008vz}. 

It is worth noting that $a-c$ controls and relates to many phenomena in CFT and holography. These include the mixed current-gravitational anomaly \cite{Anselmi:1997am} in SCFTs; superconformal indices and their high temperature asymptotics \cite{DiPietro:2014bca, Ardehali:2014zba, Beccaria:2014xda}; violations of the KSS bound on $\eta/s$ in holographic CFTs \cite{Kovtun:2004de, Brigante:2007nu}; and the size of the single-trace higher spin gap in large $N$ SCFTs \cite{Camanho:2014apa}. 

Finally, it is a remarkable and still mysterious fact that nearly all ``traditional'' CFTs have $a\leq c$ rather than $a> c$. Our result may be regarded as suggesting a naturalness of such asymmetry, along the lines of \cite{DiPietro:2014bca}. It would be very interesting to make this more concrete.

\label{sec:discussion}

\begin{acknowledgments}

It is a pleasure to thank  H.~Casini, F.~Haehl, V.~Hubeny, Z.~Komargodski, A.~Lewkowycz, H.~Maxfield, M.~Mezei, S.~Solodukhin and E.~Tonni for discussions.

E.~Perlmutter and M.~Rangamani would like to thank KITP, Santa Barbara for hospitality during the concluding stages of the project, where their work was supported in part by the National Science Foundation under Grant No. NSF PHY11-25915.
M.~Rangamani was supported in part by the STFC Consolidated Grant ST/L000407/1, and by the ERC Consolidator Grant Agreement ERC-2013-CoG-615443: SPiN. E.~Perlmutter was supported by the Department of Energy under Grant No. DE-FG02-91ER40671.

\end{acknowledgments}


\begin{thebibliography}{46}
\expandafter\ifx\csname natexlab\endcsname\relax\def\natexlab#1{#1}\fi
\expandafter\ifx\csname bibnamefont\endcsname\relax
  \def\bibnamefont#1{#1}\fi
\expandafter\ifx\csname bibfnamefont\endcsname\relax
  \def\bibfnamefont#1{#1}\fi
\expandafter\ifx\csname citenamefont\endcsname\relax
  \def\citenamefont#1{#1}\fi
\expandafter\ifx\csname url\endcsname\relax
  \def\url#1{\texttt{#1}}\fi
\expandafter\ifx\csname urlprefix\endcsname\relax\def\urlprefix{URL }\fi
\providecommand{\bibinfo}[2]{#2}
\providecommand{\eprint}[2][]{\url{#2}}

\bibitem[{\citenamefont{Solodukhin}(2008)}]{Solodukhin:2008dh}
\bibinfo{author}{\bibfnamefont{S.~N.} \bibnamefont{Solodukhin}},
  \bibinfo{journal}{Phys.Lett.} \textbf{\bibinfo{volume}{B665}},
  \bibinfo{pages}{305} (\bibinfo{year}{2008}), \eprint{0802.3117}.

\bibitem[{\citenamefont{Myers and Sinha}(2011)}]{Myers:2010tj}
\bibinfo{author}{\bibfnamefont{R.~C.} \bibnamefont{Myers}} \bibnamefont{and}
  \bibinfo{author}{\bibfnamefont{A.}~\bibnamefont{Sinha}},
  \bibinfo{journal}{JHEP} \textbf{\bibinfo{volume}{1101}}, \bibinfo{pages}{125}
  (\bibinfo{year}{2011}), \eprint{1011.5819}.

\bibitem[{\citenamefont{Casini et~al.}(2011)\citenamefont{Casini, Huerta, and
  Myers}}]{Casini:2011kv}
\bibinfo{author}{\bibfnamefont{H.}~\bibnamefont{Casini}},
  \bibinfo{author}{\bibfnamefont{M.}~\bibnamefont{Huerta}}, \bibnamefont{and}
  \bibinfo{author}{\bibfnamefont{R.~C.} \bibnamefont{Myers}},
  \bibinfo{journal}{JHEP} \textbf{\bibinfo{volume}{1105}}, \bibinfo{pages}{036}
  (\bibinfo{year}{2011}), \eprint{1102.0440}.

\bibitem[{\citenamefont{Jafferis et~al.}(2011)\citenamefont{Jafferis, Klebanov,
  Pufu, and Safdi}}]{Jafferis:2011zi}
\bibinfo{author}{\bibfnamefont{D.~L.} \bibnamefont{Jafferis}},
  \bibinfo{author}{\bibfnamefont{I.~R.} \bibnamefont{Klebanov}},
  \bibinfo{author}{\bibfnamefont{S.~S.} \bibnamefont{Pufu}}, \bibnamefont{and}
  \bibinfo{author}{\bibfnamefont{B.~R.} \bibnamefont{Safdi}},
  \bibinfo{journal}{JHEP} \textbf{\bibinfo{volume}{1106}}, \bibinfo{pages}{102}
  (\bibinfo{year}{2011}), \eprint{1103.1181}.

\bibitem[{\citenamefont{Vidal and Werner}(2002)}]{Vidal:2002zz}
\bibinfo{author}{\bibfnamefont{G.}~\bibnamefont{Vidal}} \bibnamefont{and}
  \bibinfo{author}{\bibfnamefont{R.}~\bibnamefont{Werner}},
  \bibinfo{journal}{Phys.Rev.} \textbf{\bibinfo{volume}{A65}},
  \bibinfo{pages}{032314} (\bibinfo{year}{2002}).

\bibitem[{\citenamefont{Audenaert et~al.}(2003)\citenamefont{Audenaert, Plenio,
  and Eisert}}]{Audenaert:2003aa}
\bibinfo{author}{\bibfnamefont{K.}~\bibnamefont{Audenaert}},
  \bibinfo{author}{\bibfnamefont{M.~B.} \bibnamefont{Plenio}},
  \bibnamefont{and} \bibinfo{author}{\bibfnamefont{J.}~\bibnamefont{Eisert}},
  \bibinfo{journal}{Phys. Rev. Lett.} \textbf{\bibinfo{volume}{90}},
  \bibinfo{pages}{027901} (\bibinfo{year}{2003}),
  \urlprefix\url{http://link.aps.org/doi/10.1103/PhysRevLett.90.027901}.

\bibitem[{\citenamefont{Plenio}(2005)}]{Plenio:2005aa}
\bibinfo{author}{\bibfnamefont{M.}~\bibnamefont{Plenio}},
  \bibinfo{journal}{Phys. Rev. Lett.} \textbf{\bibinfo{volume}{95}},
  \bibinfo{pages}{090503} (\bibinfo{year}{2005}), \eprint{quant-ph/0505071},
  \urlprefix\url{http://arxiv.org/abs/quant-ph/0505071}.

\bibitem[{\citenamefont{Rangamani and Rota}(2014)}]{Rangamani:2014ywa}
\bibinfo{author}{\bibfnamefont{M.}~\bibnamefont{Rangamani}} \bibnamefont{and}
  \bibinfo{author}{\bibfnamefont{M.}~\bibnamefont{Rota}},
  \bibinfo{journal}{JHEP} \textbf{\bibinfo{volume}{1410}}, \bibinfo{pages}{60}
  (\bibinfo{year}{2014}), \eprint{1406.6989}.

\bibitem[{\citenamefont{Calabrese et~al.}(2012)\citenamefont{Calabrese, Cardy,
  and Tonni}}]{Calabrese:2012ew}
\bibinfo{author}{\bibfnamefont{P.}~\bibnamefont{Calabrese}},
  \bibinfo{author}{\bibfnamefont{J.}~\bibnamefont{Cardy}}, \bibnamefont{and}
  \bibinfo{author}{\bibfnamefont{E.}~\bibnamefont{Tonni}},
  \bibinfo{journal}{Phys.Rev.Lett.} \textbf{\bibinfo{volume}{109}},
  \bibinfo{pages}{130502} (\bibinfo{year}{2012}), \eprint{1206.3092}.

\bibitem[{\citenamefont{Calabrese et~al.}(2013)\citenamefont{Calabrese, Cardy,
  and Tonni}}]{Calabrese:2012nk}
\bibinfo{author}{\bibfnamefont{P.}~\bibnamefont{Calabrese}},
  \bibinfo{author}{\bibfnamefont{J.}~\bibnamefont{Cardy}}, \bibnamefont{and}
  \bibinfo{author}{\bibfnamefont{E.}~\bibnamefont{Tonni}},
  \bibinfo{journal}{J.Stat.Mech.} \textbf{\bibinfo{volume}{1302}},
  \bibinfo{pages}{P02008} (\bibinfo{year}{2013}), \eprint{1210.5359}.

\bibitem[{\citenamefont{Kulaxizi et~al.}(2014)\citenamefont{Kulaxizi,
  Parnachev, and Policastro}}]{Kulaxizi:2014nma}
\bibinfo{author}{\bibfnamefont{M.}~\bibnamefont{Kulaxizi}},
  \bibinfo{author}{\bibfnamefont{A.}~\bibnamefont{Parnachev}},
  \bibnamefont{and}
  \bibinfo{author}{\bibfnamefont{G.}~\bibnamefont{Policastro}},
  \bibinfo{journal}{JHEP} \textbf{\bibinfo{volume}{1409}}, \bibinfo{pages}{010}
  (\bibinfo{year}{2014}), \eprint{1407.0324}.

\bibitem[{\citenamefont{Calabrese et~al.}(2015)\citenamefont{Calabrese, Cardy,
  and Tonni}}]{Calabrese:2014yza}
\bibinfo{author}{\bibfnamefont{P.}~\bibnamefont{Calabrese}},
  \bibinfo{author}{\bibfnamefont{J.}~\bibnamefont{Cardy}}, \bibnamefont{and}
  \bibinfo{author}{\bibfnamefont{E.}~\bibnamefont{Tonni}},
  \bibinfo{journal}{J.Phys.} \textbf{\bibinfo{volume}{A48}},
  \bibinfo{pages}{015006} (\bibinfo{year}{2015}), \eprint{1408.3043}.

\bibitem[{\citenamefont{Ryu and Takayanagi}(2006)}]{Ryu:2006ef}
\bibinfo{author}{\bibfnamefont{S.}~\bibnamefont{Ryu}} \bibnamefont{and}
  \bibinfo{author}{\bibfnamefont{T.}~\bibnamefont{Takayanagi}},
  \bibinfo{journal}{JHEP} \textbf{\bibinfo{volume}{0608}}, \bibinfo{pages}{045}
  (\bibinfo{year}{2006}), \eprint{hep-th/0605073}.

\bibitem[{\citenamefont{Allais and Mezei}(2015)}]{Allais:2014ata}
\bibinfo{author}{\bibfnamefont{A.}~\bibnamefont{Allais}} \bibnamefont{and}
  \bibinfo{author}{\bibfnamefont{M.}~\bibnamefont{Mezei}},
  \bibinfo{journal}{Phys.Rev.} \textbf{\bibinfo{volume}{D91}},
  \bibinfo{pages}{046002} (\bibinfo{year}{2015}), \eprint{1407.7249}.

\bibitem[{\citenamefont{Mezei}(2015)}]{Mezei:2014zla}
\bibinfo{author}{\bibfnamefont{M.}~\bibnamefont{Mezei}},
  \bibinfo{journal}{Phys.Rev.} \textbf{\bibinfo{volume}{D91}},
  \bibinfo{pages}{045038} (\bibinfo{year}{2015}), \eprint{1411.7011}.

\bibitem[{\citenamefont{Astaneh et~al.}(2014)\citenamefont{Astaneh, Gibbons,
  and Solodukhin}}]{Astaneh:2014uba}
\bibinfo{author}{\bibfnamefont{A.~F.} \bibnamefont{Astaneh}},
  \bibinfo{author}{\bibfnamefont{G.}~\bibnamefont{Gibbons}}, \bibnamefont{and}
  \bibinfo{author}{\bibfnamefont{S.~N.} \bibnamefont{Solodukhin}},
  \bibinfo{journal}{Phys.Rev.} \textbf{\bibinfo{volume}{D90}},
  \bibinfo{pages}{085021} (\bibinfo{year}{2014}), \eprint{1407.4719}.

\bibitem[{\citenamefont{Willmore}(1965)}]{Willmore:1965uq}
\bibinfo{author}{\bibfnamefont{T.~J.} \bibnamefont{Willmore}},
  \bibinfo{journal}{An. St. Univ. Iasi, sIa Mat. B}
  \textbf{\bibinfo{volume}{11B}}, \bibinfo{pages}{493} (\bibinfo{year}{1965}).

\bibitem[{\citenamefont{Fursaev}(2012)}]{Fursaev:2012mp}
\bibinfo{author}{\bibfnamefont{D.}~\bibnamefont{Fursaev}},
  \bibinfo{journal}{JHEP} \textbf{\bibinfo{volume}{1205}}, \bibinfo{pages}{080}
  (\bibinfo{year}{2012}), \eprint{1201.1702}.

\bibitem[{\citenamefont{Hofman and Maldacena}(2008)}]{Hofman:2008ar}
\bibinfo{author}{\bibfnamefont{D.~M.} \bibnamefont{Hofman}} \bibnamefont{and}
  \bibinfo{author}{\bibfnamefont{J.}~\bibnamefont{Maldacena}},
  \bibinfo{journal}{JHEP} \textbf{\bibinfo{volume}{0805}}, \bibinfo{pages}{012}
  (\bibinfo{year}{2008}), \eprint{0803.1467}.

\bibitem[{\citenamefont{Lee et~al.}(2014)\citenamefont{Lee, McGough, and
  Safdi}}]{Lee:2014xwa}
\bibinfo{author}{\bibfnamefont{J.}~\bibnamefont{Lee}},
  \bibinfo{author}{\bibfnamefont{L.}~\bibnamefont{McGough}}, \bibnamefont{and}
  \bibinfo{author}{\bibfnamefont{B.~R.} \bibnamefont{Safdi}},
  \bibinfo{journal}{Phys.Rev.} \textbf{\bibinfo{volume}{D89}},
  \bibinfo{pages}{125016} (\bibinfo{year}{2014}), \eprint{1403.1580}.

\bibitem[{\citenamefont{Lewkowycz and Perlmutter}(2015)}]{Lewkowycz:2014jia}
\bibinfo{author}{\bibfnamefont{A.}~\bibnamefont{Lewkowycz}} \bibnamefont{and}
  \bibinfo{author}{\bibfnamefont{E.}~\bibnamefont{Perlmutter}},
  \bibinfo{journal}{JHEP} \textbf{\bibinfo{volume}{1501}}, \bibinfo{pages}{080}
  (\bibinfo{year}{2015}), \eprint{1407.8171}.

\bibitem[{\citenamefont{Zyczkowski}(2003)}]{zycz}
\bibinfo{author}{\bibfnamefont{K.}~\bibnamefont{Zyczkowski}},
  \bibinfo{journal}{Open Syst. Inf. Dyn.} \textbf{\bibinfo{volume}{10}},
  \bibinfo{pages}{297} (\bibinfo{year}{2003}), \eprint{0305062}.

\bibitem[{\citenamefont{{Marques} and {Neves}}(2014)}]{Marques:2014ul}
\bibinfo{author}{\bibfnamefont{F.~C.} \bibnamefont{{Marques}}}
  \bibnamefont{and} \bibinfo{author}{\bibfnamefont{A.}~\bibnamefont{{Neves}}},
  \bibinfo{journal}{Annals of Mathematics} \textbf{\bibinfo{volume}{179}},
  \bibinfo{pages}{683} (\bibinfo{year}{2014}), \eprint{1202.6036}.

\bibitem[{\citenamefont{Lawson}(1970)}]{Lawson:1970bh}
\bibinfo{author}{\bibfnamefont{H.~B.} \bibnamefont{Lawson}},
  \bibinfo{journal}{Annals of Mathematics} pp. \bibinfo{pages}{335--374}
  (\bibinfo{year}{1970}).

\bibitem[{\citenamefont{Kusner}(1989)}]{kusner1989}
\bibinfo{author}{\bibfnamefont{R.}~\bibnamefont{Kusner}},
  \bibinfo{journal}{Pacific J. Math.} \textbf{\bibinfo{volume}{138}},
  \bibinfo{pages}{317} (\bibinfo{year}{1989}),
  \urlprefix\url{http://projecteuclid.org/euclid.pjm/1102650153}.

\bibitem[{\citenamefont{Simon}(1986)}]{simon}
\bibinfo{author}{\bibfnamefont{L.}~\bibnamefont{Simon}}, in
  \emph{\bibinfo{booktitle}{Miniconference on Geometry and Partial Differential
  Equations}} (\bibinfo{publisher}{Centre for Mathematical Analysis, The
  Australian National University}, \bibinfo{address}{Canberra AUS},
  \bibinfo{year}{1986}), pp. \bibinfo{pages}{187--216},
  \urlprefix\url{http://projecteuclid.org/euclid.pcma/1416336690}.

\bibitem[{\citenamefont{Joshi and S\'equin}(2007)}]{Joshi07c.:energy}
\bibinfo{author}{\bibfnamefont{P.}~\bibnamefont{Joshi}} \bibnamefont{and}
  \bibinfo{author}{\bibfnamefont{C.}~\bibnamefont{S\'equin}},
  \bibinfo{journal}{Computer-Aided Design and Applications}
  \textbf{\bibinfo{volume}{4}}, \bibinfo{pages}{1} (\bibinfo{year}{2007}),
  \urlprefix\url{http://citeseerx.ist.psu.edu/viewdoc/summary?doi=10.1.1.93.3465}.

\bibitem[{\citenamefont{Heller and Schmitt}(2015)}]{heller}
\bibinfo{author}{\bibfnamefont{S.}~\bibnamefont{Heller}} \bibnamefont{and}
  \bibinfo{author}{\bibfnamefont{N.}~\bibnamefont{Schmitt}},
  \bibinfo{journal}{Experimental Mathematics} \textbf{\bibinfo{volume}{24}},
  \bibinfo{pages}{65} (\bibinfo{year}{2015}),
  \urlprefix\url{http://dx.doi.org/10.1080/10586458.2014.954294}.

\bibitem[{\citenamefont{Brendle}(2013)}]{Brendle}
\bibinfo{author}{\bibfnamefont{S.}~\bibnamefont{Brendle}},
  \bibinfo{journal}{Bull. Math. Sciences} \textbf{\bibinfo{volume}{3}},
  \bibinfo{pages}{133} (\bibinfo{year}{2013}), \eprint{1307.6938}.

\bibitem[{\citenamefont{Kuhnel and Pinkall}(1986)}]{kuhnelpink}
\bibinfo{author}{\bibfnamefont{W.}~\bibnamefont{Kuhnel}} \bibnamefont{and}
  \bibinfo{author}{\bibfnamefont{U.}~\bibnamefont{Pinkall}},
  \bibinfo{journal}{Quart. J. Math. Oxford} \textbf{\bibinfo{volume}{37}},
  \bibinfo{pages}{437} (\bibinfo{year}{1986}),
  \urlprefix\url{http://qjmath.oxfordjournals.org/content/37/4/437.full.pdf}.

\bibitem[{\citenamefont{Hung et~al.}(2011)\citenamefont{Hung, Myers, Smolkin,
  and Yale}}]{Hung:2011nu}
\bibinfo{author}{\bibfnamefont{L.-Y.} \bibnamefont{Hung}},
  \bibinfo{author}{\bibfnamefont{R.~C.} \bibnamefont{Myers}},
  \bibinfo{author}{\bibfnamefont{M.}~\bibnamefont{Smolkin}}, \bibnamefont{and}
  \bibinfo{author}{\bibfnamefont{A.}~\bibnamefont{Yale}},
  \bibinfo{journal}{JHEP} \textbf{\bibinfo{volume}{1112}}, \bibinfo{pages}{047}
  (\bibinfo{year}{2011}), \eprint{1110.1084}.

\bibitem[{\citenamefont{Donnelly and Wall}(2015)}]{Donnelly:2014fua}
\bibinfo{author}{\bibfnamefont{W.}~\bibnamefont{Donnelly}} \bibnamefont{and}
  \bibinfo{author}{\bibfnamefont{A.~C.} \bibnamefont{Wall}},
  \bibinfo{journal}{Phys.Rev.Lett.} \textbf{\bibinfo{volume}{114}},
  \bibinfo{pages}{111603} (\bibinfo{year}{2015}), \eprint{1412.1895}.

\bibitem[{\citenamefont{Huang}(2014)}]{Huang:2014pfa}
\bibinfo{author}{\bibfnamefont{K.-W.} \bibnamefont{Huang}}
  (\bibinfo{year}{2014}), \eprint{1412.2730}.

\bibitem[{\citenamefont{Dowker}(2010)}]{Dowker:2010bu}
\bibinfo{author}{\bibfnamefont{J.}~\bibnamefont{Dowker}}
  (\bibinfo{year}{2010}), \eprint{1009.3854}.

\bibitem[{\citenamefont{Intriligator et~al.}(1995)\citenamefont{Intriligator,
  Seiberg, and Shenker}}]{Intriligator:1994rx}
\bibinfo{author}{\bibfnamefont{K.~A.} \bibnamefont{Intriligator}},
  \bibinfo{author}{\bibfnamefont{N.}~\bibnamefont{Seiberg}}, \bibnamefont{and}
  \bibinfo{author}{\bibfnamefont{S.}~\bibnamefont{Shenker}},
  \bibinfo{journal}{Phys.Lett.} \textbf{\bibinfo{volume}{B342}},
  \bibinfo{pages}{152} (\bibinfo{year}{1995}), \eprint{hep-ph/9410203}.

\bibitem[{\citenamefont{Gaiotto}(2012)}]{Gaiotto:2009we}
\bibinfo{author}{\bibfnamefont{D.}~\bibnamefont{Gaiotto}},
  \bibinfo{journal}{JHEP} \textbf{\bibinfo{volume}{1208}}, \bibinfo{pages}{034}
  (\bibinfo{year}{2012}), \eprint{0904.2715}.

\bibitem[{\citenamefont{Bah et~al.}(2012)\citenamefont{Bah, Beem, Bobev, and
  Wecht}}]{Bah:2012dg}
\bibinfo{author}{\bibfnamefont{I.}~\bibnamefont{Bah}},
  \bibinfo{author}{\bibfnamefont{C.}~\bibnamefont{Beem}},
  \bibinfo{author}{\bibfnamefont{N.}~\bibnamefont{Bobev}}, \bibnamefont{and}
  \bibinfo{author}{\bibfnamefont{B.}~\bibnamefont{Wecht}},
  \bibinfo{journal}{JHEP} \textbf{\bibinfo{volume}{1206}}, \bibinfo{pages}{005}
  (\bibinfo{year}{2012}), \eprint{1203.0303}.

\bibitem[{\citenamefont{Gaiotto and Maldacena}(2012)}]{Gaiotto:2009gz}
\bibinfo{author}{\bibfnamefont{D.}~\bibnamefont{Gaiotto}} \bibnamefont{and}
  \bibinfo{author}{\bibfnamefont{J.}~\bibnamefont{Maldacena}},
  \bibinfo{journal}{JHEP} \textbf{\bibinfo{volume}{1210}}, \bibinfo{pages}{189}
  (\bibinfo{year}{2012}), \eprint{0904.4466}.

\bibitem[{\citenamefont{Buchel et~al.}(2009)\citenamefont{Buchel, Myers, and
  Sinha}}]{Buchel:2008vz}
\bibinfo{author}{\bibfnamefont{A.}~\bibnamefont{Buchel}},
  \bibinfo{author}{\bibfnamefont{R.~C.} \bibnamefont{Myers}}, \bibnamefont{and}
  \bibinfo{author}{\bibfnamefont{A.}~\bibnamefont{Sinha}},
  \bibinfo{journal}{JHEP} \textbf{\bibinfo{volume}{0903}}, \bibinfo{pages}{084}
  (\bibinfo{year}{2009}), \eprint{0812.2521}.

\bibitem[{\citenamefont{Anselmi et~al.}(1998)\citenamefont{Anselmi, Freedman,
  Grisaru, and Johansen}}]{Anselmi:1997am}
\bibinfo{author}{\bibfnamefont{D.}~\bibnamefont{Anselmi}},
  \bibinfo{author}{\bibfnamefont{D.}~\bibnamefont{Freedman}},
  \bibinfo{author}{\bibfnamefont{M.~T.} \bibnamefont{Grisaru}},
  \bibnamefont{and} \bibinfo{author}{\bibfnamefont{A.}~\bibnamefont{Johansen}},
  \bibinfo{journal}{Nucl.Phys.} \textbf{\bibinfo{volume}{B526}},
  \bibinfo{pages}{543} (\bibinfo{year}{1998}), \eprint{hep-th/9708042}.

\bibitem[{\citenamefont{Di~Pietro and Komargodski}(2014)}]{DiPietro:2014bca}
\bibinfo{author}{\bibfnamefont{L.}~\bibnamefont{Di~Pietro}} \bibnamefont{and}
  \bibinfo{author}{\bibfnamefont{Z.}~\bibnamefont{Komargodski}},
  \bibinfo{journal}{JHEP} \textbf{\bibinfo{volume}{1412}}, \bibinfo{pages}{031}
  (\bibinfo{year}{2014}), \eprint{1407.6061}.

\bibitem[{\citenamefont{Ardehali et~al.}(2014)\citenamefont{Ardehali, Liu, and
  Szepietowski}}]{Ardehali:2014zba}
\bibinfo{author}{\bibfnamefont{A.~A.} \bibnamefont{Ardehali}},
  \bibinfo{author}{\bibfnamefont{J.~T.} \bibnamefont{Liu}}, \bibnamefont{and}
  \bibinfo{author}{\bibfnamefont{P.}~\bibnamefont{Szepietowski}},
  \bibinfo{journal}{JHEP} \textbf{\bibinfo{volume}{1412}}, \bibinfo{pages}{145}
  (\bibinfo{year}{2014}), \eprint{1407.6024}.

\bibitem[{\citenamefont{Beccaria and Tseytlin}(2014)}]{Beccaria:2014xda}
\bibinfo{author}{\bibfnamefont{M.}~\bibnamefont{Beccaria}} \bibnamefont{and}
  \bibinfo{author}{\bibfnamefont{A.~A.} \bibnamefont{Tseytlin}},
  \bibinfo{journal}{JHEP} \textbf{\bibinfo{volume}{1411}}, \bibinfo{pages}{114}
  (\bibinfo{year}{2014}), \eprint{1410.3273}.

\bibitem[{\citenamefont{Kovtun et~al.}(2005)\citenamefont{Kovtun, Son, and
  Starinets}}]{Kovtun:2004de}
\bibinfo{author}{\bibfnamefont{P.}~\bibnamefont{Kovtun}},
  \bibinfo{author}{\bibfnamefont{D.~T.} \bibnamefont{Son}}, \bibnamefont{and}
  \bibinfo{author}{\bibfnamefont{A.~O.} \bibnamefont{Starinets}},
  \bibinfo{journal}{Phys.Rev.Lett.} \textbf{\bibinfo{volume}{94}},
  \bibinfo{pages}{111601} (\bibinfo{year}{2005}), \eprint{hep-th/0405231}.

\bibitem[{\citenamefont{Brigante et~al.}(2008)\citenamefont{Brigante, Liu,
  Myers, Shenker, and Yaida}}]{Brigante:2007nu}
\bibinfo{author}{\bibfnamefont{M.}~\bibnamefont{Brigante}},
  \bibinfo{author}{\bibfnamefont{H.}~\bibnamefont{Liu}},
  \bibinfo{author}{\bibfnamefont{R.~C.} \bibnamefont{Myers}},
  \bibinfo{author}{\bibfnamefont{S.}~\bibnamefont{Shenker}}, \bibnamefont{and}
  \bibinfo{author}{\bibfnamefont{S.}~\bibnamefont{Yaida}},
  \bibinfo{journal}{Phys.Rev.} \textbf{\bibinfo{volume}{D77}},
  \bibinfo{pages}{126006} (\bibinfo{year}{2008}), \eprint{0712.0805}.

\bibitem[{\citenamefont{Camanho et~al.}(2014)\citenamefont{Camanho, Edelstein,
  Maldacena, and Zhiboedov}}]{Camanho:2014apa}
\bibinfo{author}{\bibfnamefont{X.~O.} \bibnamefont{Camanho}},
  \bibinfo{author}{\bibfnamefont{J.~D.} \bibnamefont{Edelstein}},
  \bibinfo{author}{\bibfnamefont{J.}~\bibnamefont{Maldacena}},
  \bibnamefont{and} \bibinfo{author}{\bibfnamefont{A.}~\bibnamefont{Zhiboedov}}
  (\bibinfo{year}{2014}), \eprint{1407.5597}.

\end{thebibliography}

\end{document}